\documentclass[preprint]{revtex4}  
\usepackage{amsmath}
\usepackage{amssymb}
\usepackage{amsbsy}
\usepackage{graphicx}
\usepackage{wasysym}
\usepackage[dvips]{epsfig}
\usepackage{psfrag}
\usepackage[ansinew]{inputenc}
\usepackage[T1]{fontenc}
\usepackage{lmodern}
\usepackage{xcolor}

\newcommand{\ds}{\displaystyle}
\newcommand{\ul}{\underline}

\newcommand{\p}{\partial}

\newcommand{\epsi}{\varepsilon}

\newcommand{\bra}[1]{\left\langle{#1}\right|}

\newcommand{\braket}[2]{\left\langle{#1} \,|\, {#2} \right\rangle}

\newcommand{\op}[1]{\mathsf #1}
\newcommand{\set}[1]{\mathcal{#1}}

\renewcommand{\vec}[1]{\mbox{\boldmath $ #1 $}}

\newcommand{\N}{\mathbb{N}}

\def\gu{\;{\lower0.3ex\hbox{$\buildrel > \over {\scriptstyle \sim}$}}\;}
\def\lu{\;{\lower0.3ex\hbox{$\buildrel < \over {\scriptstyle \sim}$}}\;}

\def\XXint#1#2#3{{\setbox0=\hbox{$#1{#2#3}{\int}$}
     \vcenter{\hbox{$#2#3$}}\kern-.5\wd0}}

\begin{document}

\title{Kinetic damping in the spectra of the\\ spherical impedance probe}
\author{J.~Oberrath}
\affiliation{Leuphana University L\"uneburg,\\Institute of Product and Process Innovation,\\ Volgershall 1, 21339 L\"uneburg, Germany}
\date{\today}
\begin{abstract}

The impedance probe is a measurement device to measure plasma parameter like electron density. It consists of one electrode connected to a network analyzer via a coaxial cable and is immersed into a plasma. A bias potential superposed with an alternating potential is applied to the electrode and the response of the plasma is measured. Its dynamical interaction with the plasma in electrostatic, kinetic description can be modeled in an abstract notation based on functional analytic methods. These methods provide the opportunity to derive a general solution, which is given as the response function of the probe-plasma system. It is defined by the matrix elements of the resolvent of an appropriate dynamical operator. Based on the general solution a residual damping for vanishing pressure can be predicted and can only be explained by kinetic effects. Within this manuscript an explicit response function of the spherical impedance probe is derived. Therefore, the resolvent is determined by its algebraic representation based on an expansion in orthogonal basis functions. This allows to compute an approximated response function and its corresponding spectra. These spectra show additional damping due to kinetic effects and are in good agreement with former kinetically determined spectra.

\end{abstract}
\maketitle

\section{Introduction}
First investigations of the so called \textit{resonance probe} (RP) go back to the year 1960 and were done by Takayama, Ikegami, and Miyazaki \cite{takayama1960}. They applied a negative bias potential superposed with a small alternating potential at a planar disk electrode immersed into a plasma and measured the current. Sweeping the frequency of the alternating potential, they observed a resonance phenomenon close to the electron plasma frequency $\omega_{\rm p}$. Further theoretical and experimental works showed, that the resonance frequency $\omega_{\rm r}$ has to be below the electron plasma frequency \cite{levitskii1963,harp1964}.

In subsequent years the RP was intensively investigated, experimentally and theoretically. Many researchers have made attempts at this task, especially at the resonance probe with a spherical electrode \cite{fejer1964, crawford1964, dote1965, lepechinsky1966, buckley1966, balmain1966, davis1966, buckley1967, tarstrup1972, aso1973, bantin1974, kist1977}. The theoretical works of the cited papers have in common, that the underlying models are based on an electrostatic approximation, but the plasma description is of different complexity. They range from cold and warm fluid models to kinetic descriptions. 

An extensive kinetic analysis was done by Buckley \cite{buckley1966}. He derived and solved an integral equation of the alternating electric field to determine the impedance and the admittance as response functions of the excited plasma. Calculated and measured spectra of the admittance were in good agreement with measurements for certain moderate collision frequencies $\nu_0$ \cite{davis1966}. Beside the collisional damping the spectra showed also collisionles damping due to kinetic effects, which was found to be dominant for $\nu_0\leq 0.1\,\omega_{\rm p}$ \cite{buckley1967}. However, the physical mechanism of the kinetic damping was not explained. 

Several years later Morin and Balmain compared the kinetic admittances of Buckley with admittances determined by a fluid model \cite{morin1991}. The resonance frequencies were in good agreement for both: a continuous equilibrium density profile and a single step density profile. A meaningful difference was observed in the half width of the resonance peaks due to kinetic damping, but they also did not explain its physical mechanism. 

Within the last decade the spherical RP gained new interest and is now called spherical impedance probe (sIP). Two slightly different designs are proposed, analyzed, and characterized \cite{blackwell2005, hopkins2014}. The sIP is also discussed in the context of industry compatible plasma diagnostics \cite{kim2016}, but is still not a standard tool in industry. One possible reason for that might be the kinetic influence on the spectra, which is not fully understood, yet.    

\pagebreak 

It is also of interest to study generic features of such probes independently of any particular realization. Using methods of functional analysis, a general investigation of such probes in electrostatic approximation is given in \cite{lapke2013}. Based on the cold plasma model the main result was that, for any possible probe design, the spectral response function could be expressed as a matrix element of the resolvent of the dynamical operator. 

A fully kinetic generalization of the study of \cite{lapke2013},
i.e., an abstract kinetic model of electrostatic resonance valid for all pressures is presented in \cite{oberrath2014}. It turned out that the main result could directly be transferred. In particular, it still holds that, for any possible probe design, 
the spectral response of the probe-plasma system can be expressed as a matrix element of the resolvent of the dynamical operator. Furthermore, it was shown that the corresponding resonances exhibit a residual damping in the limit of vanishing pressure which cannot be explained by Ohmic dissipation but only by kinetic effects. The analysis of the abstract model allowed to interpret this kinetic damping as loss of kinetic free energy $\mathfrak F$. The free energy is produced by the probe, transported through the plasma to a large distance, where the probe is unable to detect it. This loss of free energy is recorded in the spectrum of the probe as damping.  

Functional analytic methods are very useful to study generic features, but they can also be applied to determine explicit spectra for a specific probe design. In spherical geometry it is possible to solve the analytic solution of the fluiddynamical response, which was applied to the idealized sIP and the multipole resonance probe, respectively \cite{oberrath2014b}. In the kinetic description an analytic solution is impossible, but an efficient algorithm can be derived to determine an adequate approximation. Such an algorithm is applied to the parallel electrode probe (PEP) \cite{oberrath2016}, which is not used for real measurements. It is meant as a toymodel, because it provides the simplest available geometry. The calculated spectra of the PEP show kinetic damping as predicted by the general analysis.  

However, the predicted kinetic damping from the general model in functional analytic description is not verified for an existing probe design, yet. Within this manuscript we focus on the idealized sIP and determine its approximated response function by means of functional analytic methods. Therefore, we follow the approximation algorithm presented in \cite{oberrath2016} and compare the spectra with the spectra of Buckley \cite{buckley1967}. It will be shown, that both are in good agreement: the spectra of the admittance and impedance.   

\pagebreak

\section{Model of the idealized spherical impedance probe}\label{sec:model}  
\begin{figure}[h!]
\vspace{-7mm}\hspace{-2cm}
\includegraphics[width=0.5\columnwidth]{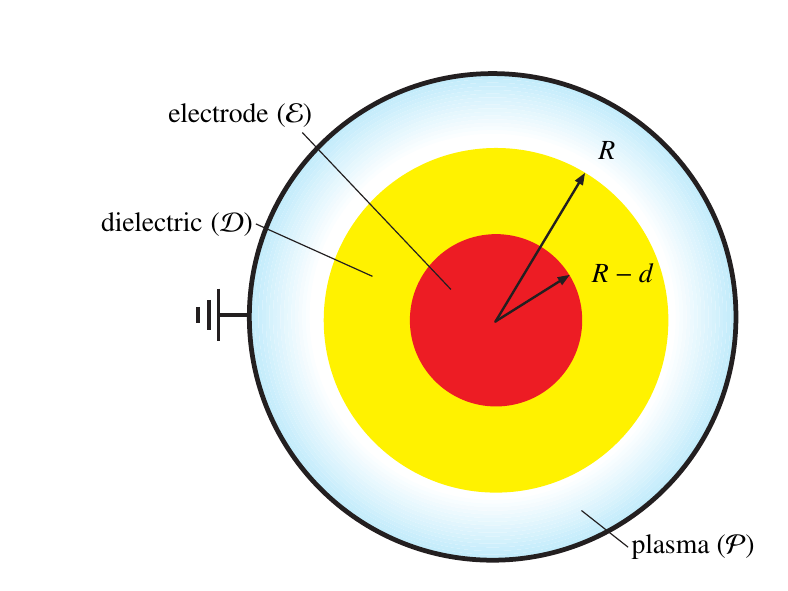} 
\vspace{-5mm}
\caption{Illustration of the idealized spherical impedance probe with the powered electrode ${\set E}$, the dielectric $\set D$, and the perturbed plasma $\set P$. The radius of the electrode is $R-d$ and the thickness of the dielectric is $d$.}
\label{idealIPkin}
\end{figure}

As depicted in fig.~\ref{idealIPkin} the idealized sIP consists of one spherical electrode $\set E$ of radius $R-d$. In a general case it can be surrounded by a dielectric $\set D$ of thickness $d$. Applying an RF voltage $U$ at the electrode, the plasma will be dynamically disturbed in the surrounding of the probe. Then $\set P$ is called the disturbed plasma and $\set V=\set P\cup\set D$ the influence domain of the probe. The former interface $\set F$ between the perturbed and unperturbed plasma, which was presented in the general model of APRS \cite{oberrath2014}, is treated as a grounded spherical surface at a large distance $R_{\infty}$ -- theoretically in an infinite distance. 

The dynamical behavior of the probe-plasma system in $\set P$ can be described by the linearized and normalized Boltzmann equation in electrostatic approximation. In the geometry of the sIP the 6 dimensional distribution function reduces to 3 dimensions due to symmetry. It depends on the radial distance $r\in[R,R_{\infty}]$, the absolute value of the velocity $v\in[0,\infty)$, and the projection angle $\chi\in[0,\pi]$ of $\vec v$ to the $r$ direction and is given by
\begin{align} 
\frac{\p g}{\p t}
&+ v\cos(\chi)\left(\frac{\p g}{\p r}-\frac{\p \op\Phi}{\p r}\right) 
 + \frac{\p \bar\Phi}{\p r}
   \left(\cos(\chi)\frac{\p g}{\p v}
   -\frac{\sin(\chi)}{v}\frac{\p g}{\p \chi}\right)
  \nonumber\\
&- U v\cos(\chi)\frac{\p \psi}{\p r} 
 =\frac{\nu_0}{2}\int_0^{\pi} g \sin(\chi)\, d\chi -\nu_0  g \ .
\label{LinBE} 
\end{align}

The perturbed distribution function $g$ of the electrons is defined in $\set P$ with homogeneous boundary conditions at the surface of the probe $g(R,v,\chi , t)=0$ and the outer grounded surface $g(R_{\infty},v,\chi ,t)=0$. Pure elastic collisions with a constant collision frequency $\nu_0$ are taken into account between electrons and the neutral background. 
 
$\op\Phi$ is called inner potential. It is a linear functional of $g$ and obeys Poisson's equation
\begin{equation}
\frac{1}{r^2}\frac{\p}{\p r}\left(r^2\varepsilon(r)\frac{\p\op\Phi}{\p r}\right)=
\left\{\begin{matrix}
0 & r\in\set D\\[1ex]
\ds 2\pi\int_{0}^{\pi} \int_{0}^{\infty} w\, g \sin(\chi)\,dv\, d\chi  & r\in\set P
\end{matrix}\right.
\label{PoissonIP}
\end{equation}
with homogeneous boundary conditions $\op\Phi(R-d)=\op\Phi(R_{\infty})=0$. $\varepsilon(r)$ is the dielectric constant and defined as $\varepsilon(r)=\varepsilon_D$ in $\set D$ and $\varepsilon(r)=1$ in $\set P$. In addition, $w$ is a positive weighting function. It is defined as the negative derivative of the equilibrium distribution $F(\epsilon)$ with respect to the total energy $\epsilon=\frac{1}{2}v^2-\bar\Phi$ in equilibrium. Assuming a Maxwellian distribution, $w$ is equal to
\begin{equation}
w(r,v)=\frac{1}{(2\pi)^{\frac{3}{2}}}e^{-\frac{v^2}{2}+\bar\Phi(r)}
\end{equation} 
where $\bar\Phi$ is 
the potential 
in equilibrium.

The radio frequent excitation of the probe is represented by the electrode function $\psi(r)$. It follows Laplace's equation
\begin{equation}
\frac{1}{r^2}\frac{\p}{\p r}\left(r^2\frac{\p\psi^{(\set{D}/\set{P})}}{\p r}\right) = 0\label{Laplace}
\end{equation}
and fulfills the boundary conditions $\psi^{(\set D)}(R-d)=1$ and $\psi^{(\set D)}(R_{\infty})=0$ and the transition conditions $\psi^{(\set{P})}(R)=\psi^{(\set{D})}(R)$ and $\psi'\, ^{(\set{P})}(R)=\varepsilon_D\psi'\,^{(\set{D})}(R)$. The solutions are easily determined
\begin{eqnarray}
\psi^{(\set{D})}(r) & = & 
\frac{(d-R) (r (R_{\infty} (\varepsilon_D-1)-R \epsilon_D)+R R_{\infty})}
     {r \varepsilon_D (d-R) (R_{\infty}-R)-d rR_{\infty}}\ ,\\[1ex]
\psi^{(\set{P})}(r) & = & \frac{R \varepsilon_D (R-d) (r-R_{\infty})}
     {r \varepsilon_D (d-R) (R_{\infty}-R)-d rR_{\infty}}
     \label{charIP}\ .
\end{eqnarray}

\section{Inner admittance in functional analytic description}

Since the model of the idealized sIP is defined, the results of the general analysis presented in \cite{oberrath2014} can be applied. Thus, the complete current $I$ at the electrode is given by the sum of the vacuum current $i_{\rm vac}$, which is present even without plasma, and the inner current $i_{\rm in}$:
\begin{equation}
I=i_{\rm vac}+i_{\rm in} =\left(Y_{\rm vac}+Y\right)\, U=Y_{\rm IP}\, U\ .
\end{equation}
The vacuum admittance $Y_{\rm vac}$ is determined by the characteristic function $\psi(r)$ and is defined as
\begin{equation}
Y_{\rm vac}=-4 \pi \varepsilon_D (R-d)^2\,i\omega\left.\frac{\p}{\p r}\psi^{(\set{D})}\right|_{r=R-d}\ .
\end{equation}
$i_{\rm in}$ can be written in a Hilbert space notation as 
\begin{equation}
i_{\rm in}=\bra{e}\left(i\omega-\op T_V-\op T_S\right)^{-1} e\,\rangle\, U
   =Y\, U\ .
\label{current}
\end{equation}
The inner admittance $Y$ is determined by the scalar product of the excitation and observation vector $e=v\cos(\chi)\frac{\p}{\p r}\psi$ and the resolvent of the dynamical operator $\op{T}_V+\op{T}_S$
\begin{equation}
Y=\langle e |(i\omega-\op{T}_V-\op{T}_S)^{-1} e\rangle \ .
\label{responseIP}
\end{equation}
$\op T_V$ and $\op T_S$ are the Vlasov and the collision operator, respectively. They, applied to a dynamical state vector $g$, are defined as follows:
\begin{eqnarray}
\op T_V g & = &  v\cos(\chi)\left(\frac{\p \op\Phi}{\p r}-\frac{\p g}{\p r}\right) 
+\frac{\p \bar\Phi}{\p r}
 \left(\frac{\sin(\chi)}{v}\frac{\p g}{\p \chi}-\cos(\chi)\frac{\p g}{\p v}\right)
 \label{VlasovOp}\ ,\\[1ex]
\op T_S g & = & \frac{\nu_0}{2}\int_0^{\pi} g \sin(\chi)\, d\chi -\nu_0  g  \ .  
\end{eqnarray}
Of particular importance in the functional analytic description is the scalar product, which is used in \eqref{responseIP}. To allow for physical interpretations it has to be connected to the system dynamics and is motivated by the kinetic free energy $\mathfrak{F}$. Generally, the scalar product of two dynamical state vectors $g'$ and $g$ is given by
\begin{align}
\langle g^\prime | g \rangle 
&=\braket{g'}{g}_{\set P}+\braket{g'}{g}_{\set V}\label{ScalarProduct}\\[1ex] 
&=2\sqrt{2\pi}\int_{R}^{R_\infty}\int_0^\pi\int_0^\infty  g^{\prime*} g\, n\,e^{-\frac{v^2}{2}} \sin(\chi)v^2 r^2\, dv\, d\chi\, dr
+4\pi\int_{R-d}^{R_\infty}\varepsilon\frac{\p \mathsf\Phi^{\prime*}}{\p r}\frac{\p \mathsf\Phi}{\p r} r^2\,dr\nonumber\ . 
\end{align}
Based on this scalar product $Y$ can be expanded by means of a complete orthonormal basis $\{a\}$ of the Hilbert space. Introducing the corresponding completeness relation twice into equation \eqref{responseIP} yields
\begin{equation}
Y
=\sum_{a'}\braket{e}{a'}\sum_{a}\langle a' |
\left(i\omega-\op{T}_V-\op{T}_S\right)^{-1} a\rangle\braket{a}{e}\ .
\label{responseExp}
\end{equation}
This equates to a vector-matrix-vector multiplication which is determined by the algebraic representation of the resolvent. The algebraic representation of the resolvent can be calculated by the inverse of the algebraic representation of $i\omega-\op T_V-\op T_S$ \cite{oberrath2014b}.

\pagebreak

\section{Explicit expansion of the inner admittance}

In the previous section the general expansion of $Y$ is shown. Now, one can follow the algorithm presented in \cite{oberrath2016} to determine an explicit expansion for the sIP with finite dimension
\begin{equation}
Y
=\vec e^{T}\cdot\left(i\omega\ul{\op I}-\ul{\op{T}}_{\,V}-\ul{\op{T}}_{\,S}\right)^{-1}\cdot\vec e\ .
\label{responseExpSimp}
\end{equation}
The three matrices in \eqref{responseExpSimp} are: the identity matrix $\ul{\op I}$, the collision matrix $\ul{\op{T}}_{\,S}$, and the Vlasov matrix $\ul{\op{T}}_{\,V}$. $\vec e$ is the explicitly expanded excitation vector. 

To determine the identity matrix, a complete orthonormal basis is needed. An appropriate basis function in velocity space is given by
\begin{equation}
g_{v}^{\kappa\lambda}(v,\chi)
=\pi^{\frac{1}{4}}\Lambda_{\kappa}^\lambda(v)\bar P_\lambda(\cos(\chi))\label{BasisKetv}\ .
\end{equation}
$\lambda\in\mathbb N_0$ is the expansion index for the projection angle and $\kappa\in\mathbb N_0$ for the absolute value of the velocity. $\bar P_\lambda(\cos(\chi))$ are the normalized Legendre polynomials, which are orthonormal and complete on the interval $\chi\in[0,\pi]$
\begin{equation}
\bar P_\lambda(\chi)
=\sqrt{\frac{2\lambda+1}{2}}P_\lambda(\cos(\chi))\ .
\end{equation}
$\Lambda_{\kappa}^\lambda(v)$ are based on the generalized Laguerre polynomials $L_\kappa^{\lambda+\frac{1}{2}}\left(\frac{1}{2}v^2\right)$. Due to the exponential part in the weighting function $w$, they are an adequate choice on the interval $v\in[0,\infty]$. They become the orthonormal functions $\Lambda_{\kappa}^{\lambda}(v)$ 
with an additional factor  
\begin{equation}
\Lambda_{\kappa}^{\lambda}(v)
=\sqrt{\frac{\kappa !}{\Gamma(\kappa+\lambda+\frac{3}{2})}}
 \left(\frac{v^2}{2}\right)^{\frac{\lambda}{2}} 
      L_\kappa^{\lambda+\frac{1}{2}}\left(\frac{v^2}{2}\right)\ .
\end{equation}

In physical space it is difficult to determine an orthogonal function due to the two different parts of the scalar product. Therefore, the focus is placed on the first part $\braket{g'}{g}_{\set P}$ of \eqref{ScalarProduct}. An appropriate basis function can be defined as
\begin{equation}
g_r^{k}(r)=\frac{\sin\left(k\pi\frac{r-R}{R_\infty-R}\right)}{\sqrt{2\pi\,e^{\bar\Phi(r)}\,(R_\infty-R)}\, r}
\end{equation}
with $k\in\N$. It is orthonormal and complete on the interval $[R,R_\infty]$ and fulfills the boundary conditions $g_r^{k}(R)=g_r^{k}(R_\infty)=0$. In summary,  
\begin{equation}
g_{k}^{\kappa\lambda}(r,v,\chi)
=g_r^{k}(r)g_v^{\kappa\lambda}(v,\chi)\label{BasisKet}
\end{equation} 
is an orthonormal and complete basis function based on the scalar product $\braket{g'}{g}_{\set P}$. 

\pagebreak 
Indeed, this basis function is nether orthonormal nor complete based on the complete scalar product \eqref{ScalarProduct}. However, it can be used to determine a non-diagonal basis matrix $\ul{\op B}$, which can be diagonalized afterwords. To do so, the complete scalar product of two basis functions has to be computed and thus the derivative of the inner potential is needed. It is given by 
\begin{equation}
\frac{\p\op\Phi_k^{\lambda\kappa}}{\p r}
=\frac{\delta _{\kappa 0}\, \delta_{\lambda 0}}{r^2}\left\{\begin{matrix}
A_k^{(\set D)} & r\in\set D\\[1ex]
\ds A_k^{(\set P)}
   +\int_R^{r} {r''}^2 \, e^{\bar\Phi(r)}\, g_r^{k} \, dr'' \, dr'  & r\in\set P
\end{matrix}\right.
\ .
\label{SimpGradPotIP}
\end{equation} 
($\delta _{\kappa 0}$ and $\delta_{\lambda 0}$ are Kronecker deltas. The derivation is presented in the Appendix \ref{sec:Pot}). Equation \eqref{SimpGradPotIP} shows, that the inner potential $\op\Phi_k^{\lambda\kappa}$ of a basis function is zero for all $\lambda\neq0\neq\kappa$. The same holds for the inner potential $\op\Phi_{k'}^{\lambda'\kappa'}$ if $\lambda'\neq0\neq\kappa'$. Thus, the second part of the scalar product simplifies to 
\begin{equation}
\braket{g_{k'}^{\kappa'\lambda'}}{g_{k}^{\kappa\lambda}}_{\set V}
=4\pi\int_{R-d}^{R_\infty}\varepsilon\frac{\p \mathsf\Phi_{k'}^{\lambda'\kappa'}}{\p r}\frac{\p \mathsf\Phi_k^{\lambda\kappa}}{\p r} r^2\,dr\,\delta_{\kappa 0}\delta_{\lambda 0}\delta_{\kappa' 0}\delta_{\lambda' 0}=\op B_{kk'}^{(00)} 
\end{equation} 
and is not zero only if $\lambda=\lambda'=\kappa=\kappa'=0$. This leads to two different results for the complete scalar product of the basis functions
\begin{eqnarray}
\braket{g_{k'}^{00}}{g_{k}^{00}} & = & \op B_{kk'}^{(00)}+\delta_{kk'}\label{basisB0}\ ,\\
\braket{g_{k'}^{\kappa'\lambda'}}{g_{k}^{\kappa\lambda}} & = & \delta_{kk'}\delta_{\kappa\kappa'}\delta_{\lambda\lambda'}\ .
\end{eqnarray}
These expressions are the elements of the basis matrix $\ul{\op B}$, which is a block diagonal matrix. It is almost diagonal. The only non-diagonal block is the first block $\ul{\op B}^{(00)}$ with the elements in equation \eqref{basisB0}. This block can be diagonalized with a rotation matrix $\ul{\op C}$ to find the diagonal matrix $\ul{\op D}^{(00)}=\ul{\op C}\,\ul{\op B}^{(00)}\, \ul{\op C}^T$. Multiplying this diagonalized matrix with its inverse leads to the identity matrix $\ul{\op I}^{(00)}=\ul{\op D}^{(00)}\, {\ul{\op D}^{(00)}}^{-1}$. Then, $\ul{\op B}$ turns into a pure identity matrix $\ul{\op I}$.  

Applying the collision operator to the basis function and computing the scalar product leads to the matrix elements of the collision matrix $\ul{\op T}_{\,S}$
\begin{equation}
\langle g_{k'}^{\kappa'\lambda'} | \op T_S g_{k}^{\kappa\lambda}\rangle
=\langle g_{k'}^{\kappa'\lambda'} | \op T_S g_{k}^{\kappa\lambda}\rangle_{\set P}
=\, \nu_0 \,\left(\delta_{\lambda0}\,\delta_{0\lambda'}-\delta_{\lambda\lambda'}\right) 
  \,\delta_{\kappa\kappa'} \label{MeTs}\ .
\end{equation}
$\ul{\op T}_{\,S}$ is a diagonal matrix with zero elements on the main diagonal if $\lambda=\lambda'=0$. Due to that, no diagonalization is needed. 

The computation of the Vlasov matrix $\ul{\op T}_{\,V}$ is more complicated. Its elements are determined by the scalar product between the basis functions and the Vlasov operator $\langle g_{k'}^{\kappa'\lambda'} | \op T_V g_{k}^{\kappa\lambda}\rangle$. As shown in \cite{oberrath2016}, $\ul{\op T}_{\,V}$ is an anti-symmetric block matrix. Due to that only two inner block matrices with the indices $\kappa=\kappa'=\lambda=0$, $\lambda'=1$ and $\kappa=\kappa'=\lambda'=0$, $\lambda=1$ have to be multiplied with the rotation matrices $\ul{\op C}$ or $\ul{\op C}^T$ to get the correct expanded Vlasov matrix. Detailed calculations to the Vlasov matrix can be found in the appendix \ref{sec:VlasovMatrix}. 

Finally, the excitation vector $\vec e$ has to be determined. Its elements are defined by
\begin{equation}
\braket{g_{k'}^{\kappa'\lambda'}}{e}
=
4\pi\int_R^{R_\infty} r^2 \, e^{\bar\Phi(r)}\, g_r^{k'}\, \frac{\p\psi}{\p r} \,dr
\,\delta_{\lambda' 1}\,\delta_{\kappa' 0}=e_{k'}\,\delta_{\lambda' 1}\,\delta_{\kappa' 0}\ .
\end{equation}
Obviously, $\vec e$ has non-vanishing elements only for $\kappa'=0$ and $\lambda'=1$ and is given by 
\begin{equation}
\vec e=\left(\vec 0\, ,\, {\ul{\op D}^{(00)}}^{-\frac{1}{2}}\,\ul{\op C}\,\vec e_{k'}\, ,\, \vec 0\, ,\, \dots \right)^T\ .
\end{equation}
Since all matrices and the excitation vector in equation \eqref{responseExpSimp} are defined, the explicit expansion of the admittance can be calculated to compute different spectra of the sIP.

\section{Spectra of the spherical impedance probe}

Within the last section an explicit expansion of the inner admittance of the idealized sIP is derived and can be used to compute approximated spectra. To compare the first calculated kinetic spectra based on functional analytic methods for a real probe design all parameters are taken from Buckley \cite{buckley1967}. His spectra are calculated for a sIP without dielectric ($d=0$, $\varepsilon_D=1$) and a probe radius of $R=5.15\,\lambda_D$. He used the equilibrium potential $\bar\Phi(r)$ of a spherical electrode in a plasma presented by Bernstein and Rabinowitz \cite{bernstein1959}. Buckley varied the collision frequency $\nu_0\in\{0.05 ,0.15, 0.25, 0.4\}\, \omega_{\rm p}$ and normalized the admittance to $4\pi\varepsilon_0 \omega_{\rm p}R$. The distance to the outer grounded surface is chosen to be $R_\infty=150\lambda_D$, where also the plasma frequency is nomalized to $\omega_{\rm p}$. 

\begin{figure}[h!]
\centering
\includegraphics[height=4.7cm]{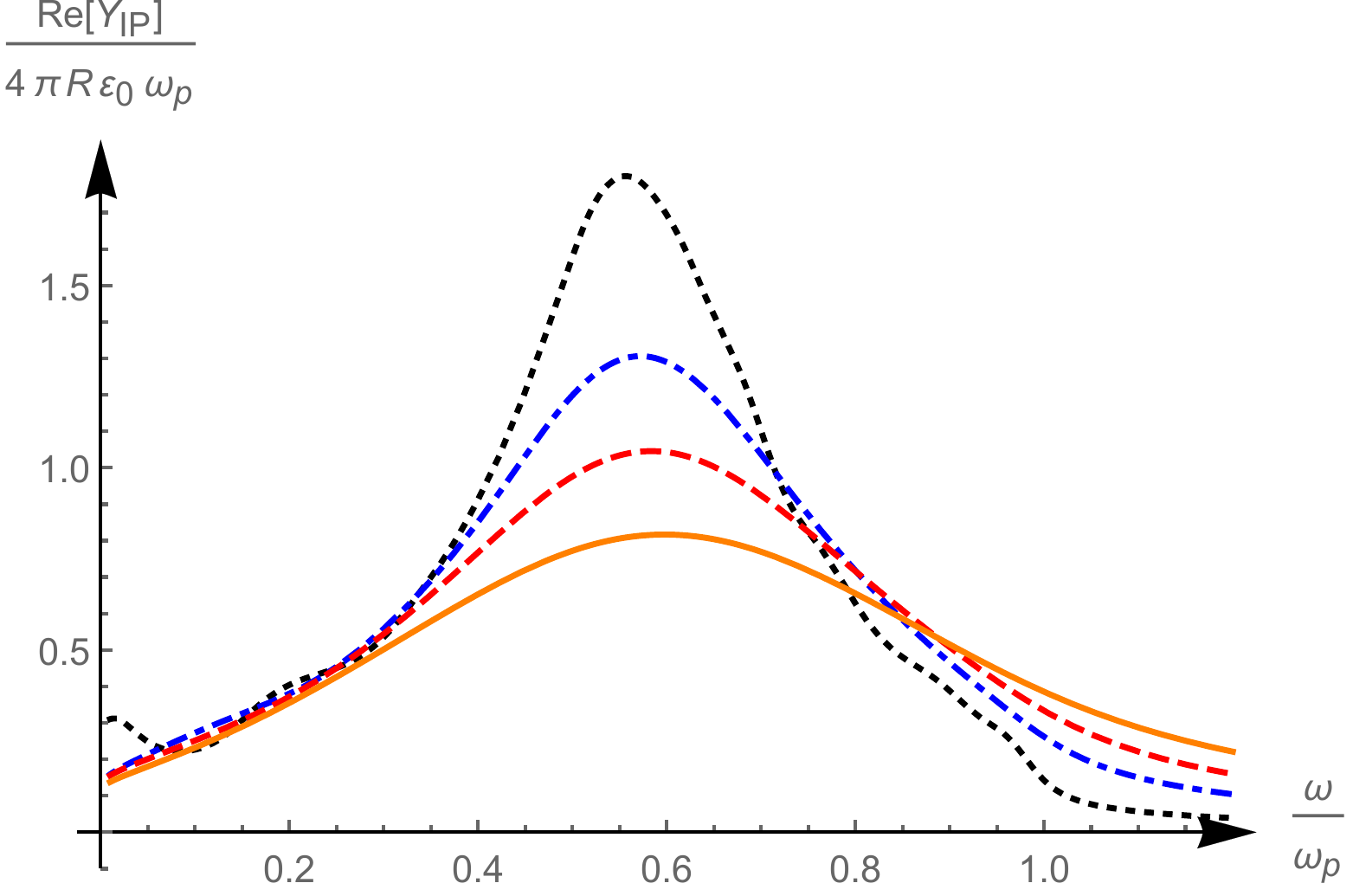}
\hspace{8mm}
\includegraphics[height=4.7cm]{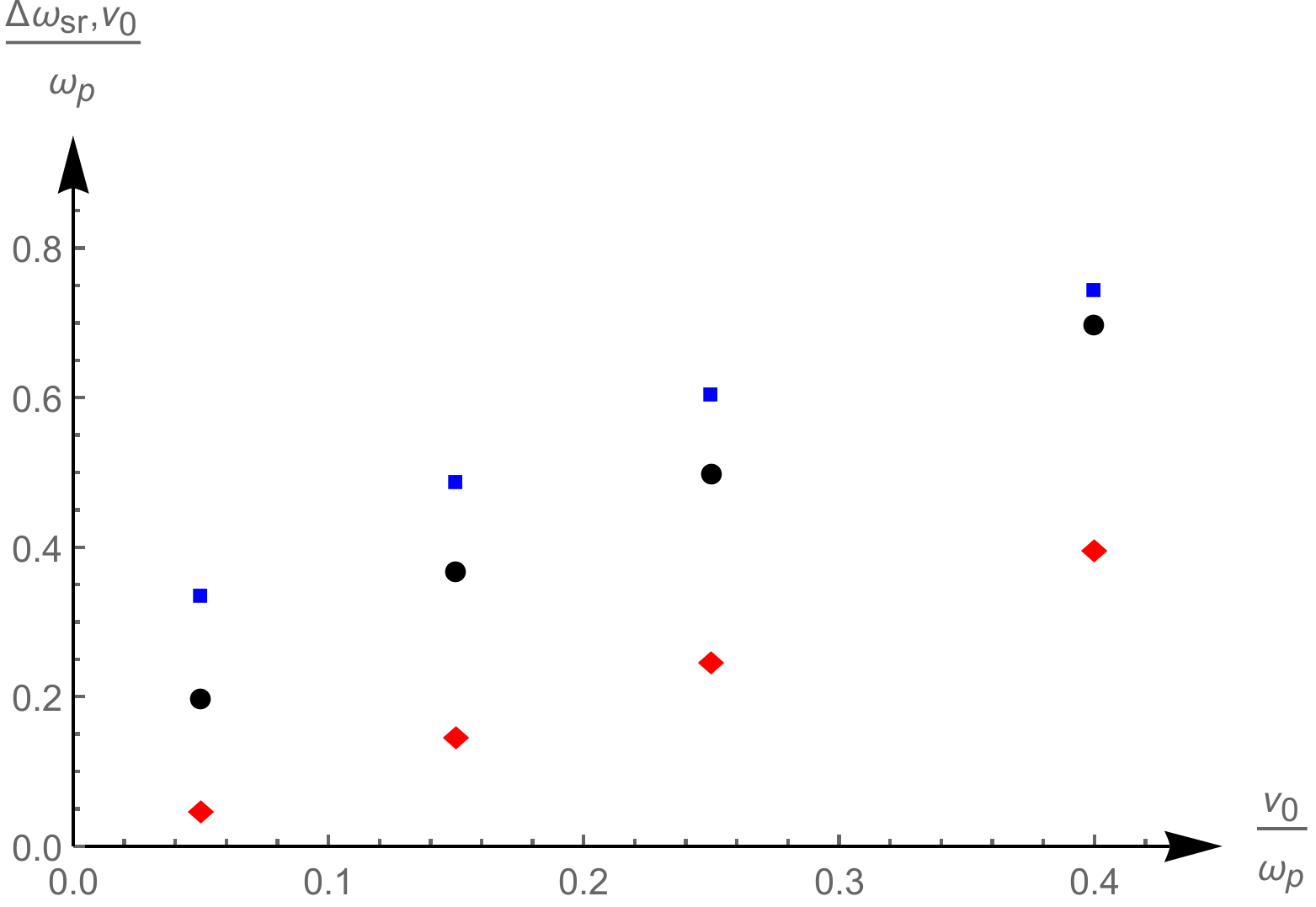}
\caption{Normalized real part of the admittance (left) of the sIP for $\kappa_{\rm max}=\lambda_{\rm max}=25$, $k_{\rm max}=500$ and different collision frequencies $\nu_0\omega_{\rm p}^{-1}$: 0.05 (dotted), \textcolor{blue}{0.15~(dot-dashed)}, \textcolor{red}{0.25~(dashed)}, and \textcolor{orange}{0.4~(bold)}. Half width of the admittance spectra (right) from Buckley (dots) and \textcolor{blue}{functional analytic calculation (squares)} and the \textcolor{red}{collision frequency (diamonds)}.}
\label{YIP}
\end{figure}

In fig. \ref{YIP} (left) the real part of the admittance $Y_{\rm IP}$ is depicted for the maximum expansion indices in velocity space $\kappa_{\rm max}=\lambda_{\rm max}=25$ and a maximum expansion index $k_{\rm max}=500$ in physical space. All spectra are almost converged and in good agreement with the spectra of Buckley, but the resonance frequencies and the half widths are not identical. Buckley's resonance frequencies increase from $\omega_{\rm srB}=0.59\, \omega_{\rm p}$ to $\omega_{\rm srB}=0.62\, \omega_{\rm p}$ by increasing the collision frequency. The resonance frequencies calculated in this manuscript also increase $\omega_{\rm sr}\in\{0.557,0.572,0.584, 0.598\}\, \omega_{\rm p}$, but are smaller than Buckley's. The half width $\Delta\omega_{\rm sr}$ from the functional analytic calculations is broader and increases less than Buckley's as shown in fig. \ref{YIP} (right). 
\pagebreak

\begin{figure}[h!]
\centering
\includegraphics[height=4.65cm]{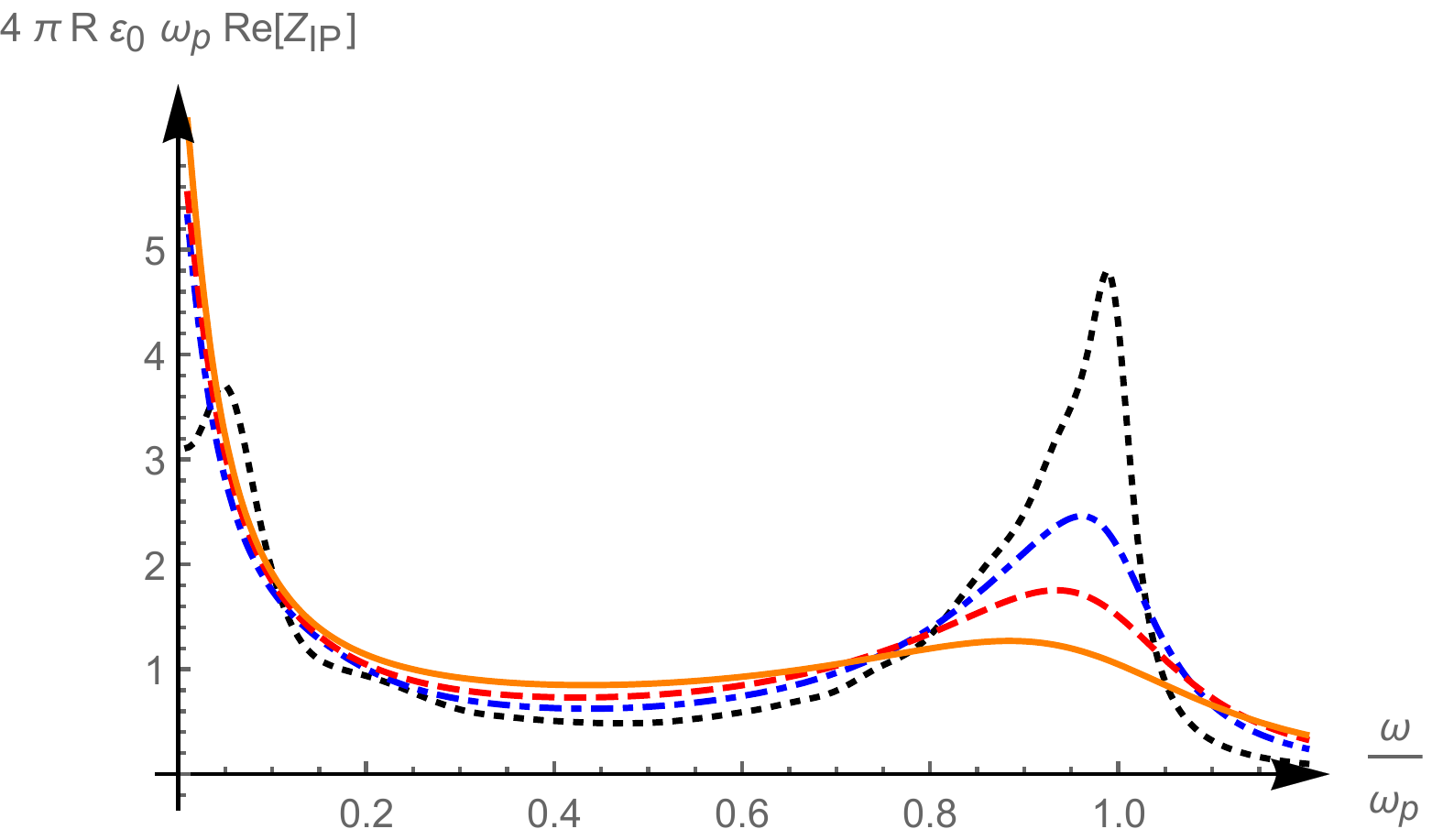}
\hspace{8mm}
\includegraphics[height=4.65cm]{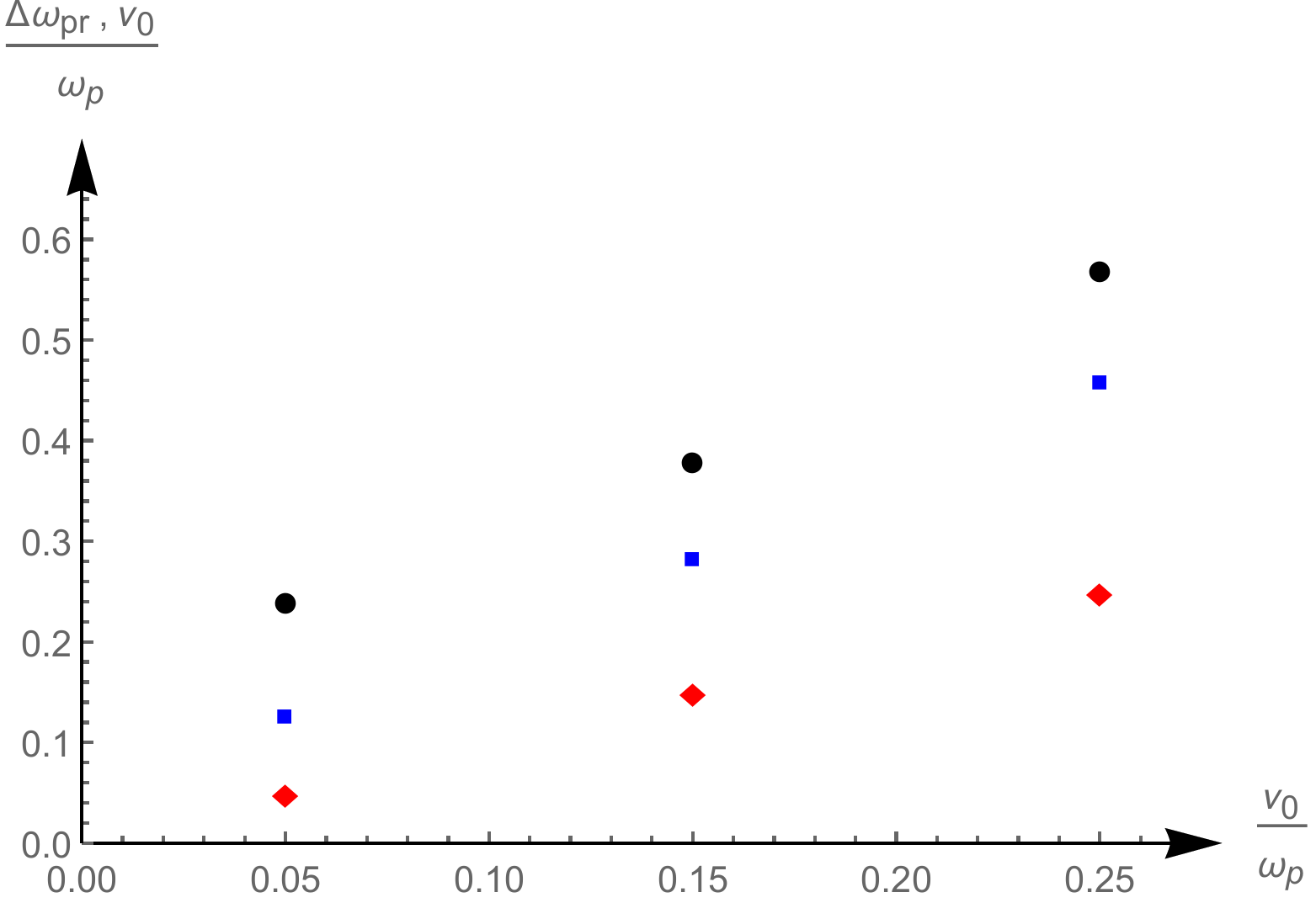}
\caption{Normalized real part of the impedance (left) of the sIP for $\kappa_{\rm max}=\lambda_{\rm max}=25$, $k_{\rm max}=500$ and different collision frequencies $\nu_0\omega_{\rm p}^{-1}$: 0.05 (dotted), \textcolor{blue}{0.15~(dot-dashed)}, \textcolor{red}{0.25~(dashed)}, and \textcolor{orange}{0.4~(bold)}. Half width of the admittance spectra (right) from Buckley (dots) and \textcolor{blue}{functional analytic calculation (squares)} and the \textcolor{red}{collision frequency (diamonds)}.}
\label{ZIP}
\end{figure}

Another resonance behavior can be observed in the spectra of the impedance $Z_{\rm IP}=Y_{\rm IP}^{-1}$, which are shown in fig. \ref{ZIP} (left). A clear resonance close to -- but smaller than -- the plasma frequency appears. Its frequency decreases with the increase of the collision frequency\linebreak ($\omega_{\rm pr}\in\{0.989, 0.961, 0.934, 0.884\}\, \omega_{\rm p}$). These resonance frequencies are larger than Buckley's, which decrease from $\omega_{\rm prB}=0.9\, \omega_{\rm p}$ to $\omega_{\rm prB}=0.8\, \omega_{\rm p}$. The half widths in the impedance spectra increase by increasing the collision frequency (see fig. \ref{ZIP} (right)). Thus, the behavior is identical to Buckley's results, but the half widths are smaller. The half width for the largest collision frequency could not be determined, due to the strong damping.    

\pagebreak
The half widths in the spectra represent the damping of the probe-plasma system. In both cases -- admittance and impedance -- they are larger than determined by fluid models, which is caused by kinetic effects. Assuming $\Delta\omega=\nu_0+\nu_{\rm kin}$, the kinetic damping can be determined. In the spectra of the admittance it is about $\nu_{\rm sr,kin}\in\{0.283, 0.336, 0.353, 0.343\}\, \omega_{\rm p}$ and in the spectra of the impedance about $\nu_{\rm pr,kin}\in\{0.074, 0.132, 0.206\}\, \omega_{\rm p}$.

\section{Conclusion}

Within this manuscript a kinetic model of the spherical impedance probe is presented. Its dynamical interaction with the plasma is given by the inner admittance of the probe-plasma system, which is determined by the resolvent of the dynamical operator $\op T_V + \op T_S$. The expanded inner admittance of the sIP is derived by means of a complete basis in its particular geometry. This leads to the matrix representation of the dynamical operator. The truncation of the expansion allows to approximate the inner admittance and thus to analyze the kinetic damping within its spectra.  

To compare the approximated spectra of the functional analytic approach, the parameters in the calculations are taken from Buckley \cite{buckley1967}. The spectra for the four different collision frequencies $\nu_0\in\{0.05,0.15,0.25,0.4\}\,\omega_{\rm p}$ are in good agreement in the position and in the height of the peaks. The resonance frequencies and the half widths are not identical, which is probably due to the different collision terms. Within this manuscript the collision term for pure elastic collisions between electrons and neutral atoms is used, which is based on a more physically reasonable derivation than the simple relaxation term used by Buckley.  

Besides the admittance spectra also impedance spectra are shown, where a clear resonance smaller than the plasma frequency can be observed. Compared to Buckley, the resonance frequencies differ slightly, but the half widths and thus the heights of the peaks differ more, which can also be explained by the different collision terms.            

All spectra show larger half widths as spectra which would be calculated by a fluid model. This additional damping is caused by kinetic effects. Due to the scalar product \eqref{ScalarProduct}, which is motivated by the kinetic free energy $\mathfrak{F}$, the kinetic loss mechanism can be interpreted: the probe produces kinetic free energy, which is transported through the plasma and escapes at a large distance of the probe, where the probe can not detect it anymore \cite{oberrath2014}. This loss of kinetic free energy is recorded in the spectrum of the probe as damping. 

An additional loss of kinetic free energy is present in the collision process itself. The collision term in the linearized Boltzmann equation \eqref{LinBE} drives the perturbed distribution function to an isotropic one. This means to lose information about the velocity direction. Information loss is equal to an increase of the kinetic entropy $\mathfrak{S}$ and thus, a decrease of the kinetic free energy $\mathfrak{F}=\mathfrak{U}-\mathfrak{S}$, where $\mathfrak{U}$ is the total energy. 

In summary it is shown that the approximated spectra based on the functional analytic approach are in good agreement with former determined spectra of Buckley. They show kinetic damping which can be explained by loss of kinetic free energy. This loss of kinetic free energy is recorded by the probe as damping. On one hand it is caused by the increase of the kinetic entropy due to elastic collisions and on the other hand by the escaped free energy to an unobservable distance to the probe. The latter is connected to the collisionless damping. 

Of course, electron depletion caused by a probe, which is immersed into a plasma, has to be taken into account in the investigations of APRS probes as stated in a recent published paper \cite{sternberg2017}. First work in that direction was already done: Bernstein and Rabinowitz \cite{bernstein1959} derived the potential of a spherical electrode surrounded by plasma, which was used by Buckley \cite{buckley1966} as equilibrium potential. Morin and Balmain \cite{morin1991} compared Buckley's results with a warm fluid model, where the resonance frequencies were in good agreement. Within their calculations, they used a continuous electron density profile in equilibrium including electron depletion from Allen et al. \cite{allen1957}. Furthermore, they compared their results with a simplified model including a single step electron density profile and observed just a small difference in the resonance frequency. Thus, a few works about the influence of the equilibrium density profile in the surrounding of a spherical probe were presented and the resonance frequencies are comparable. 

However, the author will focus on detailed investigations of the influence of pressure and temperature dependent equilibrium density profiles on the spectra of APRS probes. This is necessary on one hand to demonstrate the correctness of APRS measurements and on the other hand to derive a relation between the half width of resonance peaks in admittance spectra and the electron temperature.

\appendix

\section{Potential of a basis function}\label{sec:Pot}

Entering the basis function \eqref{BasisKet} into Poisson's equation \eqref{PoissonIP} the integrals over the velocity space can be solved which yields 
\begin{equation}
\frac{1}{r^2}\frac{\p}{\p r}\left(r^2\frac{\p\op\Phi_k^{\lambda\kappa}}{\p r}\right)
=\delta_{\kappa 0}\delta_{\lambda 0}\left\{\begin{matrix}
0 & r\in\set D\\[1ex]
\ds \, e^{\bar\Phi(r)}\, g_r^{k}(r)  & r\in\set P
\end{matrix}\right.
\ .
\label{SimpPoissonIP}
\end{equation} 

The potentials for the different regions (plasma $\set P$ and dielectric $\set D$) can be solved by integration and using the boundary conditions $\op\Phi_{k}^{(\set P)}(R_\infty)=0$ and $\op\Phi_{k}^{(\set D)}(R-d)=0$: 
\begin{eqnarray}
\op\Phi_{k\kappa\lambda}^{(\set P)}(r)
&=&A_k^{(\set P)} \left(\frac{1}{R_\infty}-\frac{1}{r}\right)
   \delta _{\kappa 0}\, \delta_{\lambda 0}\nonumber\\[1ex]
& &+\int_R^r \frac{1}{{r'}^2}\int_R^{r'} {r''}^2 \, e^{\bar\Phi(r)}\, g_r^{k} \, dr'' \, dr'
   \,\delta _{\kappa 0}\, \delta_{\lambda 0}\\[1ex]
& &-\int_R^{R_\infty} \frac{1}{{r'}^2}\int_R^{r'} {r''}^2 \, e^{\bar\Phi(r)}\, g_r^{k} \, dr'' \, dr'
    \,\delta _{\kappa 0}\, \delta_{\lambda 0}\nonumber \ ,\\[1ex]
\op\Phi_{k\kappa\lambda}^{(\set D)}(r)
&=&A_{k}^{(\set D)} \left(\frac{1}{R-d}-\frac{1}{r}\right)
   \delta _{\kappa 0}\, \delta_{\lambda 0} \ .  
\end{eqnarray}
The constants $A_k^{(\set P)}$ and $A_k^{(\set D)}$ are determined by the transition conditions
\begin{eqnarray}
\op\Phi_{k}^{(\set D)}(R) & = & \op\Phi_{k}^{(\set P)}(R)\ , \\[1ex]
\epsi_D \left.\frac{\p\op\Phi_{k}^{(\set D)}}{\p r}\right|_R
& = &  \left.\frac{\p\op\Phi_{k}^{(\set P)}}{\p r}\right|_R \ .
\end{eqnarray}
In the scalar product the derivative of the potential is needed and can be written as 
\begin{equation}
\frac{\p\op\Phi_k^{\lambda\kappa}}{\p r}
=\frac{\delta _{\kappa 0}\, \delta_{\lambda 0}}{r^2}\left\{\begin{matrix}
A_k^{(\set D)} & r\in\set D\\[1ex]
\ds A_k^{(\set P)}
   +\int_R^{r} {r''}^2 \, e^{\bar\Phi(r)}\, g_r^{k} \, dr'' \, dr'  & r\in\set P
\end{matrix}\right.
\ .
\end{equation}


\section{Matrix elements of the Vlasov-Operator}\label{sec:VlasovMatrix}

The Vlasov operator is defined in \eqref{VlasovOp}. Applied to the basis function $g_k^{\kappa\nu}$ yields
\begin{align}
\op T_V g_k^{\kappa\nu} 
=& v\cos(\chi)\left(\frac{\p \op\Phi}{\p r}-\frac{\p g_k^{\kappa\nu}}{\p r}\right) 
  +\frac{\p \bar\Phi}{\p r}
   \left(\frac{\sin(\chi)}{v}\frac{\p g_k^{\kappa\nu}}{\p \chi}
  -\cos(\chi)\frac{\p g_k^{\kappa\nu}}{\p v}\right)\ .  
\end{align}
In the scalar product the derivative of the potential $\op\Phi^{(\op T_V)}$ is needed, which is meant as the potential produced by the Vlasov operator applied to the basis function. In \cite{oberrath2016} is shown that this derivative is given by the electron particle flux within the plasma $\set P$ and vanishes within the dielectric $\set D$. In the geometry of the sIP one finds
\begin{equation}
\frac{\p\op\Phi_k^{(\op T_V)}}{\p r}
=-\frac{1}{\sqrt{2\pi}}\,g_r^{k}\, e^{\bar\Phi(r)}\int_{0}^\pi\int_0^{\infty } e^{-\frac{v^2}{2}}    
    g_{v}^{\kappa\lambda}\, v^3 \sin(\chi)\cos(\chi)\,dv\,d\chi\ .
\end{equation}
Due to that, the elements of the Vlasov matrix are given by
\begin{align}
\langle g_{k'}^{\kappa'\lambda'} | \op T_V | g_{k}^{\kappa\lambda}\rangle
= &\,\frac{1}{\sqrt{2\pi}}\int_{0}^\pi\int_0^{\infty} 
    e^{-\frac{v^2}{2}}\, g_{v}^{\kappa'\lambda'} 
    \left(\sin(\chi)\,\frac{\p g_{v}^{\kappa\lambda}}{\p\chi}
   -v \cos(\chi)\,\frac{\p g_{v}^{\kappa\lambda}}{\p v}\right) v\sin(\chi)
    \,dv\,d\chi\, {\op V}_{kk'}^{(1)}\nonumber\\[1ex]
  &+\frac{1}{\sqrt{2\pi}}\int_{0}^\pi\int_0^{\infty } e^{-\frac{v^2}{2}}\,  
    g_{v}^{\kappa'\lambda'}\, g_{v}^{\kappa\lambda}\, v^3 \sin(\chi)\cos(\chi)
    \,dv\,d\chi\, {\op V}_{kk'}^{(2)}\nonumber\\[1ex]
  &+\frac{1}{\sqrt{2\pi}}\int_{0}^\pi\int_0^{\infty }
    e^{-\frac{v^2}{2}}\, g_{v}^{\kappa'\lambda'} 
    \,\frac{\p g_{v}^{\kappa\lambda}}{\p\chi}\, v^3 \sin^2(\chi)
    \,dv\,d\chi\, {\op V}_{kk'}^{(3)}\nonumber\\[1ex]
  &+\frac{1}{\sqrt{2\pi}}\int_{0}^\pi\int_0^{\infty } e^{-\frac{v^2}{2}}\,  
    g_{v}^{\kappa'\lambda'}\, v^3 \sin(\chi)\cos(\chi)\,dv\,d\chi\,
    {\op V}_{kk'}^{(4)}\nonumber\\[1ex] 
  &+\frac{1}{\sqrt{2\pi}}\int_{0}^\pi\int_0^{\infty } e^{-\frac{v^2}{2}}    
    g_{v}^{\kappa\lambda}\, v^3 \sin(\chi)\cos(\chi)\,dv\,d\chi\, {\op V}_{kk'}^{(5)}
    \label{SPTv}
\end{align}
with
\begin{eqnarray}
{\op V}_{kk'}^{(1)} 
& = & 4\pi\int_R^{R_\infty} r^2\, g_r^{k'}\,\frac{\p e^{\bar\Phi(r)}}{\p r}\,g_r^{k} \, dr\ ,\label{SPTvV1}\\[1ex]
{\op V}_{kk'}^{(2)} 
& = & -4\pi\int_R^{R_\infty} r^2\, g_r^{k'}\, e^{\bar\Phi(r)} \frac{\p g_r^{k}}{\p r}\, dr\ ,\\[1ex]
{\op V}_{kk'}^{(3)}
& = & 4\pi\int_R^{R_\infty} r\,g_r^{k'}\, e^{\bar\Phi(r)}\, g_r^{k} \, dr\ ,\\[1ex]
{\op V}_{kk'}^{(4)} 
& = & 4\pi\int_R^{R_\infty} r^2\,g_r^{k'}\, e^{\bar\Phi(r)}\,\frac{\p\op\Phi_{k\kappa\lambda}^{(\set P)}}{\p r}\, dr\ ,\\[1ex]
{\op V}_{kk'}^{(5)} 
& = & -4\pi\int_R^{R_\infty} r^2 \,\frac{\p\op\Phi_{k'\kappa'\lambda'}^{(\set P)}}{\p r}\, e^{\bar\Phi(r)}\,g_r^{k}\, dr\label{SPTvV5}\ .
\end{eqnarray}
The integrals over the velocity space in \eqref{SPTv} can be solved analytically, but lead to long expressions. The integrals over the physical space in equations \eqref{SPTvV1} to \eqref{SPTvV5} have usually to be solved numerically, depending on the equilibrium potential $\bar\Phi(r)$.

The final Vlasov matrix $\ul{\op T}_{\,V}$ is an anti-symmetric block matrix, where the inner blocks are given by the matrices of the physical space ${\ul{\op V}^{(i)}}$ over the indices $k$ and $k'$. Due to the anti-symmetry, only the block matrices at the positions with the indices $\kappa=\kappa'=\lambda=0$, $\lambda'=1$ and $\kappa=\kappa'=\lambda'=0$, $\lambda=1$ have to be corrected for the complete orthonormal expansion. The correct block matrices at these positions are ${\ul{\op D}^{(00)}}^{-1/2}\,\ul{\op C} \,{\ul{\op V}^{(i)}}$ for $\kappa=\kappa'=\lambda=0$, $\lambda'=1$ and $ {\ul{\op V}^{(i)}}\,\ul{\op C}^T\, {\ul{\op D}^{(00)}}^{-1/2}$ for $\kappa=\kappa'=\lambda'=0$, $\lambda=1$. After this correction the complete Vlasov matrix can be computed as
\begin{equation}
\ul{\op T}_{\,V}=\sum_{i=1}^5 \ul{\op T}_{\,V}^{(i)}\ .
\end{equation}

\acknowledgments
The author acknowledges support by the internal funding of the Leuphana University Lüneburg and the German Research Foundation via the project OB 469/1-1. Gratitude is expressed to J.~Gong, D.-B.~Grys, M.~Lapke, M.~Oberberg, D.~Pohle, C.~Schulz, J.~Runkel, R.~Storch, T.~Styrnoll, S.~Wilczek, T.~Mussenbrock, P.~Awakowicz, T.~Musch, and I.~Rolfes, who are or were part of the MRP-Team at Ruhr University Bochum. Explicit gratitude is expressed to R.P.~Brinkmann for fruitful discussions.  



\begin{thebibliography}{10}

 

\bibitem{takayama1960}
K. Takayama, H. Ikegami, and S. Miyazaki, Phys. Rev. Let. {\bf 5}, 238 (1960).


\bibitem{levitskii1963}
S.M. Levitskii and I.P. Shashurin, Sov. Phys. Tech. Phys. {\bf 8}, 319 (1963).

 








\bibitem{harp1964} R. S. Harp, Appl. Phys. Lett. {\bf 4}, 186 (1964).

\bibitem{fejer1964} J.A. Fejer, Radio Sci. {\bf 68D}, 1171 (1964).


\bibitem{crawford1964}
R. S. Harp and F. W. Crawford, J. Appl. Phys. {\bf 35}, 3436 (1964).


\bibitem{dote1965}
T. Dote and T. Ichimiya, J. Appl. Phys. {\bf 36}, 1866 (1965).


\bibitem{lepechinsky1966} 
D. Lepechinsky, A.M. Messiaen, and P. Polland, J. Nucl. Energy, Part C Plasma Phys. {\bf 8}, 165 (1966).


\bibitem{buckley1966}
R. Buckley, Proc. Roy. Soc. {\bf 290}, 186 (1966). 

\bibitem{balmain1966}
K. G. Balmain, Radio Sci. {\bf 1}, 1 (1966).

\bibitem{davis1966}
P.G. Davis, Proc. Roy. Soc. {\bf 88}, 1019 (1966). 




\bibitem{buckley1967}
R. Buckley, J. Plasma Phys. {\bf 1}, 171 (1967).









\bibitem{tarstrup1972}
J. Tarstrup and W.J. Heikkila, Radio Sci. {\bf 7}, 493 (1972).










 



\bibitem{aso1973}
T. Aso, Radio Sci. {\bf 8}, 139 (1973).



\bibitem{bantin1974}
C. C. Bantin and K. G. Balmain, Can. J. Phys. {\bf 52}, 291 (1974).

\bibitem{kist1977}
R. Kist, Radio Sci. {\bf 12}, 921 (1977).



\bibitem{morin1991}
G.A. Morin and K.G. Balmain, Radio Sci. {\bf 26}, 459 (1991).






\bibitem{blackwell2005}
D. D. Blackwell, D. N. Walker, and W. E. Amatucci, Rev. Sci. Instrum {\bf 76}, 023503 (2005).


\bibitem{hopkins2014}
M.A. Hopkins and L.B. King, Phys. Plasmas {\bf 21}, 053501 (2014).

\bibitem{kim2016} 
D.W. Kim, S.J. You, J.H. Kim, H.Y. Chang, and W.Y. Oh,
\textit{Plasma Sources Sci. Technol.} {\bf 25}, 035026 (2016).











































\bibitem{lapke2013}
M. Lapke, J. Oberrath, T. Mussenbrock und R. P. Brinkmann, Plasma Scources Sci. Technol. {\bf 22}, 025005 (2013).

\bibitem{oberrath2014} 
J. Oberrath and R.P. Brinkmann, Plasma Sources Sci. Technol. {\bf 23}, 045006 (2014).


\bibitem{oberrath2014b} 
J. Oberrath and R.P. Brinkmann, Plasma Sources Sci. Technol. {\bf 23}, 065025 (2014).


\bibitem{oberrath2016} 
J. Oberrath and R.P. Brinkmann, Plasma Sources Sci. Technol. {\bf 25}, 065020 (2016).

\bibitem{bernstein1959} 
I.B. Bernstein and I.N. Rabinowitz, Phys. Fluid {\bf 2}, 112 (1959).

\bibitem{sternberg2017} 
N. Sternberg, V. Godyak, Phys. Plasmas {\bf 24}, 093504 (2017).

\bibitem{allen1957}
J.E. Allen, R.L.F. Boyd, and P. Reynolds, Proc. Phys. Soc. B {\bf 70 (3)}, 297 (1957)
















\end{thebibliography}
\end{document}